\documentclass[5p]{elsarticle}

\usepackage{amsmath,amssymb,bm}
\newcommand{\tr}{\mathop{\mathrm{tr}}}
\newcommand{\average}[1]{\langle#1\rangle}
\newcommand{\rank}{\mathop{\mathrm{rank}}}
\newcommand{\NBS}{N_\text{BS}}
\newcommand{\NEV}{{N}_\text{EV}}
\newcommand{\NNG}{N_\text{NG}}
\newcommand{\hphi}{\hat{\phi}}

\newcommand{\OP}{h}
\newcommand{\hA}{\hat{A}}
\newcommand{\hH}{\hat{H}}
\newcommand{\hL}{\hat{L}}
\newcommand{\hO}{\hat{\mathcal{O}}}
\newcommand{\hM}{\hat{M}}
\newcommand{\hN}{\hat{N}}
\newcommand{\hD}{\hat{D}}
\newcommand{\hP}{\hat{P}}
\newcommand{\hQ}{\hat{Q}}
\newcommand{\hT}{\hat{T}}
\newcommand{\hK}{\hat{K}}
\newcommand{\hrho}{\hat{\rho}}
\newcommand{\FE}{F}
\newcommand{\AM}{A}
\begin{document}

\author[tokyo,riken]{Tomoya Hayata}
\author[riken]{Yoshimasa Hidaka}
\address[tokyo]{Department of Physics, The University of Tokyo, Tokyo 113-0031, Japan}
\address[riken]{Theoretical Research Division, Nishina Center, RIKEN, Saitama 351-0198, Japan}
\title{Broken spacetime symmetries and elastic variables}
\begin{abstract}
We discuss spontaneous breaking of continuum symmetries, 
whose generators do explicitly depend on the spacetime coordinates. 
We clarify the relation between broken symmetries and elastic variables at both zero and finite temperatures, and/or finite densities,
and show the general counting rule that is model-independently determined by the symmetry breaking pattern.
We apply it to three intriguing examples: rotational, conformal, and gauge symmetries. 
\end{abstract}

\maketitle

\section{Introduction} 
Symmetry plays important role in modern physics.
When a global symmetry group $G$ is spontaneously broken into a subgroup $H$, the ground state infinitely degenerates. 
In other words,  a uniform change associated with the continuum symmetry,
characterized by a parameter $\pi$, does not cost energy. 
If $\pi$ is slowly varying in space, the energy will increase with $(\bm{\nabla}\pi)^2$.
Such a variable is called the elastic variable~\cite{Chaikin}. 
In the effective Lagrangian approach, $\pi$ is identified as  the coordinate of coset space $G/H$~\cite{Coleman:1969sm,Callan:1969sn,
Volkov:1973vd, Ogievetsky}.
For example,  in hadron physics, pions that mediate the strong force between nucleons
can be identified as the elastic variables. In a crystal, 
it corresponds to a displacement from the position of the atom at equilibrium. 

The elastic variables couple to the charges associated with the broken symmetries 
(broken charges) as canonical variables, 
and form gapless propagating modes called the Nambu-Goldstone (NG) modes~\cite{Nambu:1961tp,Goldstone:1961eq,Goldstone:1962es}.
Such  gapless modes play an important role at a low energy region. In particular, at low temperature, 
the number of them, their dispersions, and their damping rates are reflected in 
the equation of state, heat capacity, temperature dependence of the order parameter, and so on~\cite{Gasser:1984,Hofmann:1999,Hofmann:2001ck,Anderson:2002,Anderson:2004,Brauner:2010wm}.

In the case that the broken charge densities 
do not explicitly depend on the spacetime coordinates, 
the number of elastic variables $\NEV$ is equal to the number of broken symmetries (or broken charges) $\NBS$;
however, the number of NG modes $\NNG$ is not always equal to $\NEV$ or $\NBS$.  
The relation between $\NNG$  and $\NBS$ ($=\NEV$) is given by~\cite{Watanabe:2012hr,Hidaka:2012ym}
\begin{equation}
\NNG= \NBS- \frac{1}{2}\rank \average{[\hQ_a,\hQ_b]},
\label{eq:NGRelation}
\end{equation}
where  $\hQ_a$  are the broken charges.
The nonvanishing expectation value $\average{[\hQ_a,\hQ_b]}$ 
implies that the broken charge densities are also the elastic variables, and form canonical pairs~\cite{Nambu:2004}.
Such charge densities do not generate the independent NG modes, 
and thus the total number of NG modes is reduced by as many as the number of such pairs.
In fact, the second term in Eq.~(\ref{eq:NGRelation}) represents the number of the canonical pairs.
The NG modes associated with nonvanishing $\average{[\hQ_a,\hQ_b]}$ are classified as type-B, while the other NG modes are classified as type-A~\cite{Watanabe:2012hr}.
Their dispersions for type-A and type-B NG modes are linear and quadratic in momentum $|\bm{k}|$ unless fine-tuning~\cite{Watanabe:2012hr,Hidaka:2012ym}; therefore, type-A and type-B NG modes coincide with type-I and type-II NG modes classified by  Nielsen-Chadha~\cite{Nielsen:1975hm},
in which the dispersions of type-I and type-II NG modes are odd and even powers of $|\bm{k}|$, respectively.

On the other hand, when broken charge densities do explicitly depend on the spacetime coordinates 
(we refer to the symmetries as spacetime symmetries), the situation is totally different.
The counting rule of NG modes as in Eq.~(\ref{eq:NGRelation})  is not always valid.
A famous example is a crystalline order, 
in which space-translational and rotational symmetries are spontaneously broken~\cite{Chaikin}.
The charges associated with the rotational symmetries are the angular momenta that explicitly depend  on the space coordinates.
There appear the three gapless phonons (NG modes) accompanied by the translational symmetries in three dimensions, 
but no 
NG modes for the rotational symmetries.
To our knowledge, for spacetime symmetries, 
the general relation between broken generators, elastic variables, NG modes, and their dispersions has not been completely understood yet.
It is a great theoretical challenge to generalize the counting rule of NG modes to cover 
the spontaneous breaking of spacetime symmetries.

For Lorentz-invariant theories at zero temperature, the counting rule of NG modes for spontaneous breaking of spacetime symmetries was discussed by Low and Manoha~\cite{Low:2001bw}, in which  the condition that different broken symmetries do not  generate the independent NG modes was shown.
In the effective field theory approach, this can be understood by the so-called ``inverse Higgs mechanism''~\cite{Low:2001bw,Ivanov:1975zq,Nicolis:2013sga,Endlich:2013vfa,Brauner:2014aha}.
For the nonrelativistic systems,
Watanabe and Murayama proposed a criterion, ``Noether constraints,'' for the redundancy of the broken symmetries, 
using quantum operators and the vacuum state~\cite{Watanabe:2013iia}.
This criterion gives a sufficient condition and 
covers the redundancy of NG modes not only for spacetime symmetries but also for internal ones,
although the applicability is limited at zero temperature.

In this paper, as a first step towards constructing the counting rule of NG modes to cover the spacetime symmetries, 
we focus on the relation between elastic variables, corresponding to the flat directions of the free energy, and spontaneous breaking of spacetime symmetries
at both zero and finite temperatures, and/or densities.
We show the general counting rule of elastic variables that is model-independently determined by the symmetry breaking pattern.

\section{Spontaneous symmetry breaking and elastic variables}
\subsection{(Non)translationally invariant charge}
We consider a system whose microscopic theory has translational symmetries.
Charges of translations  for time and space are the Hamiltonian $\hH$ and the momentum ${\hP}^i$, respectively.
We use indices with capital ($A$, $B$, \dots) and small ($a$, $b$, \dots) letter for charges that are generators of $G$ and $G/H$, respectively.
We also use the hat symbol to indicate quantum operators. 
In general, charges transform under spatial translation $\hT_{\bm{x}}$ as 
\begin{equation}
\begin{split}
\hT_{\bm{x}}\hQ_A\hT_{\bm{x}}^{\dag}= {c_{A}}^B(\bm{x})\hQ_B.
\label{eq:chargeCommutation}
\end{split}
\end{equation}
The coefficients are given by ${c_{A}}^B(\bm{x})={[\exp(-i \bar{T}_k x^k)]_A}^B$, where ${[\bar{T}_k]_A}^B\equiv -i{f_{kA}}^B$,
with $[\hP_k,\hQ_A]= i{f_{kA}}^B\hQ_B$.
Einstein's convention on repeated indices is understood.
Since $\hT_{\bm{x}+\bm{x}'}=\hT_{\bm{x}'}\hT_{\bm{x}}$,
${c_{A}}^B(\bm{x}+\bm{x}')={c_{A}}^D(\bm{x}'){c_{D}}^B(\bm{x})$ and $[\bar{T}_k,\bar{T}_l]=0$ are satisfied. 
For Hermitian charges, ${f_{kA}}^B$ and ${c_{A}}^B(\bm{x})$ are real.
If $\hQ_A$ does not explicitly depend on the space coordinates, $\hQ_A$ commutes with $\hP_k$, i.e., ${f_{kA}}^B=0$, and  ${c_A}^B(\bm{x})={\delta_A}^B$.
We call the charge operator the translationally invariant charge.
Conversely, $\hQ_A$ that does explicitly depend on the space coordinates is called the nontranslationally invariant charge.
A typical example of nontranslationally invariant charges is the angular momentum $\hL_{ij}$, which transforms under Eq.~(\ref{eq:chargeCommutation}) as
\begin{equation}
\begin{split}
\hT_{\bm{x}}\hL_{ij} \hT^{\dag}_{\bm{x}} = \hL_{ij} -x_i\hP_j + x_j\hP_i.
\label{eq:TranslationLij}
\end{split}
\end{equation}

For later use, we define the expectation value of an arbitrary operator $\hO$ as
$\average{\hO}\equiv\tr \hrho_\text{eq}\hO$, with the Gibbs distribution density operator, $\hrho_\text{eq} \equiv \exp(-\beta \hH) / \tr \exp(-\beta \hH)$, where $\beta=1/T$ is the inverse temperature. For a nonzero chemical potential $\mu$, one may replace $\hH$ with $\hH-\mu \hN$, where $\hN$ is the number operator. The zero temperature theory is obtained by the $T\to0$ limit.

\subsection{Elastic variables} 
Let us  discuss how many elastic variables appear when continuum symmetries are spontaneously broken. 
For this purpose, we employ the free energy at finite $T$ (and $\mu$) 
because the elastic variables correspond to the flat directions of the free energy. 
First, we define the thermodynamic potential as
\begin{equation}
\begin{split}
  W[J]\equiv  -\frac{1}{\beta} \ln\tr\exp \left[-\beta \hH+ \beta\int d^dx \hphi_i(\bm{x})J^i(\bm{x}) \right] ,
\end{split}
\end{equation}
where $d$ represents the spatial dimension, and
$\hphi_i(\bm{x})$ are local-Hermitian operators belonging to a linear representation, which may be either elementary or composite.
At least, the set of $\hphi_i(\bm{x})$ is chosen to contain one order parameter for each broken generator.
We assume that $\hphi_i(\bm{x})$ transform as 
$
\hphi_i(\bm{x})=  \hT_{\bm{x}}\hphi_i(\bm{0})\hT_{\bm{x}}^\dag.
\label{eq:translationallyCovariant}
$
Next, the free energy $\FE[\phi]$ is given by the Legendre transformation of $W[J]$: 
\begin{equation}
\begin{split}
\FE[\phi]=  W[J]- \int d^dx J^i(\bm{x})\frac{\delta W[J]}{\delta J^i(\bm{x})} . \label{eq:effectiveAction}
\end{split}
\end{equation}
The first-variational derivative of $\FE[\phi]$ with respect to $\phi_j(\bm{y})$ at $J=0$ gives the stationary condition, $\delta\FE[\phi]/\delta\phi_j(\bm{y})=0$.
The second-variational derivative  at $J=0$ is equal to the inverse susceptibility,
\begin{equation}
\chi^{ij}(\bm{x},\bm{y}) = \left.\frac{\delta^2\FE[\phi]}{\delta \phi_i(\bm{x})\delta\phi_j(\bm{y})}\right|_{\phi=\average{\hphi}_{J=0}},
\label{eq:inverse Susceptibility}
\end{equation}
which satisfies
$
\int d^dw\chi_{ik}(\bm{x},\bm{w}) \chi^{kj}(\bm{w},\bm{y}) ={\delta_i}^j\delta^{(d)}(\bm{x}-\bm{y})
$, where $\chi_{ik}(\bm{x},\bm{w})\equiv \lim_{J\to0}\delta\average{\phi_i(\bm{x})}_{J}/\delta J^k(\bm{w})$ is the susceptibility.
The subscript $\langle \cdots\rangle_J$ denotes the thermal average in the external field $J$.
When we identify $\chi^{ij}(\bm{x},\bm{y})$ as an operator, we can write  the eigenvalue equation as
\begin{equation}
\begin{split}
\int d^dy\chi^{ij}(\bm{x};\bm{y}) \psi_j(n,\bm{k};\bm{y}) = \lambda_n(\bm{k}) \psi^i(n,\bm{k};\bm{x}),
\end{split}
\label{eigenvalue_Eq}
\end{equation}
where $\psi^i(n,\bm{k},\bm{x})=\psi_i(n,\bm{k},\bm{x})$. The wave function satisfies the normalization condition,
\begin{equation}
\int d^dx {\psi^{i*}}(n,\bm{k};\bm{x})\psi_i(m,\bm{k}';\bm{x})={\delta^{n}}_m(2\pi)^{d'}\delta^{(d')}(\bm{k}-\bm{k}'),
\label{eq:normalization}
\end{equation}
where $d'$ is the dimension of the unbroken translations as explained below.
The eigenvalue $\lambda_n(\bm{k})$ is nonnegative because of convexity of the free energy. 
We assume that the translational symmetry is not completely broken at least one direction, which may be discrete.
This is necessary to identify the eigenmode of unbroken translation,
which is characterized by translational operator $\hT_{\bm{R}}$ with a translation vector $\bm{R}$.
Under this translation, $\bm{x}\to \bm{x}+\bm{R}$, the eigenfunction satisfies $\psi_i(n, \bm{k};\bm{x}+\bm{R})=e^{i\bm{k}\cdot \bm{R}}\psi_i(n, \bm{k};\bm{x})$,
where $\bm{k}$ denotes a point in the first Brillouin zone. 
The eigenfunction of Eq.~(\ref{eigenvalue_Eq}) with zero eigenvalue at $\bm{k}=\bm{0}$ corresponds to the flat direction of the free energy.
Therefore, the number of elastic variables is given by the number of independent eigenfunctions of Eq.~(\ref{eigenvalue_Eq}) with zero eigenvalues at $\bm{k}=\bm{0}$.

Here, we consider  symmetry of the free energy
to find the eigenfunctions with the zero eigenvalue associated with spontaneous symmetry breaking.
For charges satisfying $[\hrho_\text{eq},\hQ_A]=0$, the free energy satisfies 
\begin{equation}
\begin{split}
\int d^dy\frac{\delta \FE[\phi] }{\delta\phi_j(\bm{y})} \average{\hat{\OP}_{Aj}(\bm{y})}_{J}=0,
\label{eq:STIdentity}
\end{split}
\end{equation}
where $\hat{\OP}_{Ai}(\bm{x})\equiv i[\hQ_A,\hphi_i(\bm{x})]=i{[T_A]_i}^j\hphi_j(\bm{x})$,
and the absence of quantum anomalies is assumed~\cite{WeinbergText}.
Differentiating Eq.~(\ref{eq:STIdentity}) with respect to $\phi_i(\bm{x})$, we obtain
\begin{equation}
\begin{split}
&\int d^dy\frac{\delta^2 \FE[\phi] }{\delta\phi_i(\bm{x})\delta\phi_j(\bm{y})} \average{\hat{\OP}_{Aj}(\bm{y})}_{J}\\
&\quad+\int d^dy\frac{\delta \FE[\phi] }{\delta\phi_j(\bm{y})} \frac{\delta  \average{\hat{\OP}_{Aj}(\bm{y})}_{J}}{\delta \phi_i(\bm{x})}=0.
\end{split}
\end{equation}
At the stationary point and $J=0$, the second term vanishes, and thus we obtain
\begin{equation}
\begin{split}
\int d^dy\chi^{ij}(\bm{x},\bm{y}) \OP_{Aj}(\bm{y})=0,
\label{eq:zeroEigenvalue}
\end{split}
\end{equation}
using Eq.~(\ref{eq:inverse Susceptibility}),
where $  \OP_{Ai}(\bm{x})\equiv\lim_{J\to0} \average{\hat{\OP}_{Ai}(\bm{x})}_{J}$.
When continuum symmetries are spontaneously broken,
there exist nonvanishing  order parameters $ \OP_{ai}(\bm{x})$,
which  are assumed to be linearly independent in the sense such that $c^a\OP_{ai}(\bm{x})=0$ if and only if $c^a=0$. 
The index $a$ runs from $1$ to $\NBS$.
We assume there exists no other elastic variable that is not accompanied by the spontaneous symmetry breaking.
Then, from Eq.~(\ref{eq:zeroEigenvalue}),
linear combinations of nonvanishing $\OP_{ai}(\bm{x})$ are  candidates of elastic variables.
To be the eigenfunctions of Eq.~(\ref{eigenvalue_Eq}), they should be eigenfunctions of unbroken spatial-translation.
When the translational symmetry is not broken,  for translationally invariant charges,
$\OP_{ai}(\bm{x})$ are constant, so that $\OP_{ai}(\bm{x})$ are eigenfunctions of translations with $\bm{k}=\bm{0}$, whose number coincides with $\NBS$,
and $\NEV=\NBS$.
However, for general cases, $\OP_{ai}(\bm{x})$ are not always eigenfunctions of translations.
To see this, let us consider an unbroken translation $\hT_{\bm{R}}$ under which 
the thermal state is invariant. The order parameter transforms into
\begin{equation}
\begin{split}
\OP_{ai}(\bm{x})
&=i\average{ [\hT_{\bm{R}}\hQ_a\hT^{\dag}_{\bm{R}},\hT_{\bm{R}}\hphi_i(\bm{x})\hT^{\dag}_{\bm{R}}]}\\
&=i{c_a}^b(\bm{R})\average{ [\hQ_b,\hphi_i(\bm{x}+\bm{R})]}\\
&={c_a}^b(\bm{R}) \OP_{bi}(\bm{x}+\bm{R}).
\end{split}
\end{equation}
If a linear combination of $\OP_{ai}(\bm{x})$, $f^a\OP_{ai}(\bm{x})$ is an eigenfunction of the unbroken translation with $\bm{k}=\bm{0}$, 
it satisfies~\footnote{We note that this linear combination must satisfy the normalization condition~Eq.~(\ref{eq:normalization}).}
\begin{equation}
\begin{split}
f^a\OP_{ai}(\bm{x})=f^a{c_a}^b(\bm{R}) \OP_{bi}(\bm{x}).
\end{split}
\end{equation}
In addition, with  ${\AM_a}^b(\bm{R})\equiv {\delta_a}^b-{c_a}^b(\bm{R})$,  $f^a{\AM_a}^b(\bm{R})=0$ must be satisfied for any translation vector $\bm{R}$.
Therefore, the number of independent-elastic variables is given by the  dimension of the nontrivial solutions of $f^a{\AM_a}^b(\bm{R})=0$, 
which equals the dimension of the kernel of ${\AM_b}^a(\bm{R})$,
\begin{equation}
\begin{split}
\NEV =\mathop{\mathrm{dim}}\mathop{\mathrm{ker}}\AM(\bm{R}), \label{eq:CountingElasticVariable}
\end{split}
\end{equation}
for arbitrary $\bm{R}$.
We note that the indices of ${\AM_a}^b(\bm{R})$ run for broken generators.
If the unbroken translation is continuum, the matrix $A$ is understood as 
${\AM_a}^b(\bm{\epsilon})=\epsilon^k{f_{ka}}^b$ with the infinitesimal-translation vector $\epsilon^k$.

This result does not mean that a nontranslationally invariant  charge does not generate any eigenfunctions. In the following, we show three examples in which broken nontranslationally invariant charges do or do not generate the elastic variables, depending on the pattern of the symmetry breaking.
\section{Examples}
In this section, we apply our result to  three systems with spontaneous breaking of non-translational symmetries:
rotational, conformal, and gauge symmetries.
\subsection{Nematic and smectic-A phases in a liquid crystal}
\begin{figure}
\includegraphics[scale=.28]{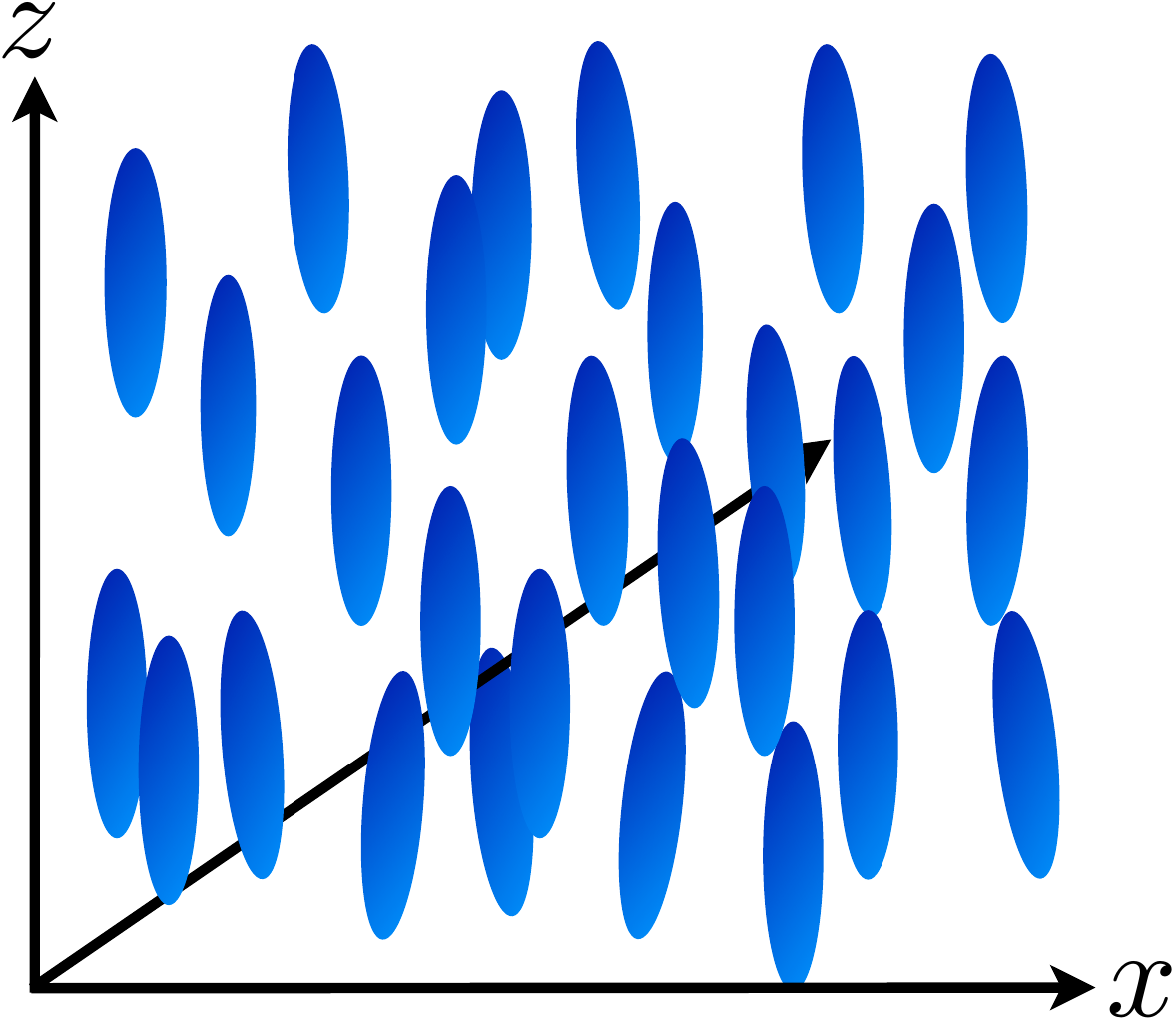}
\hspace{0.5cm}
\includegraphics[scale=.28]{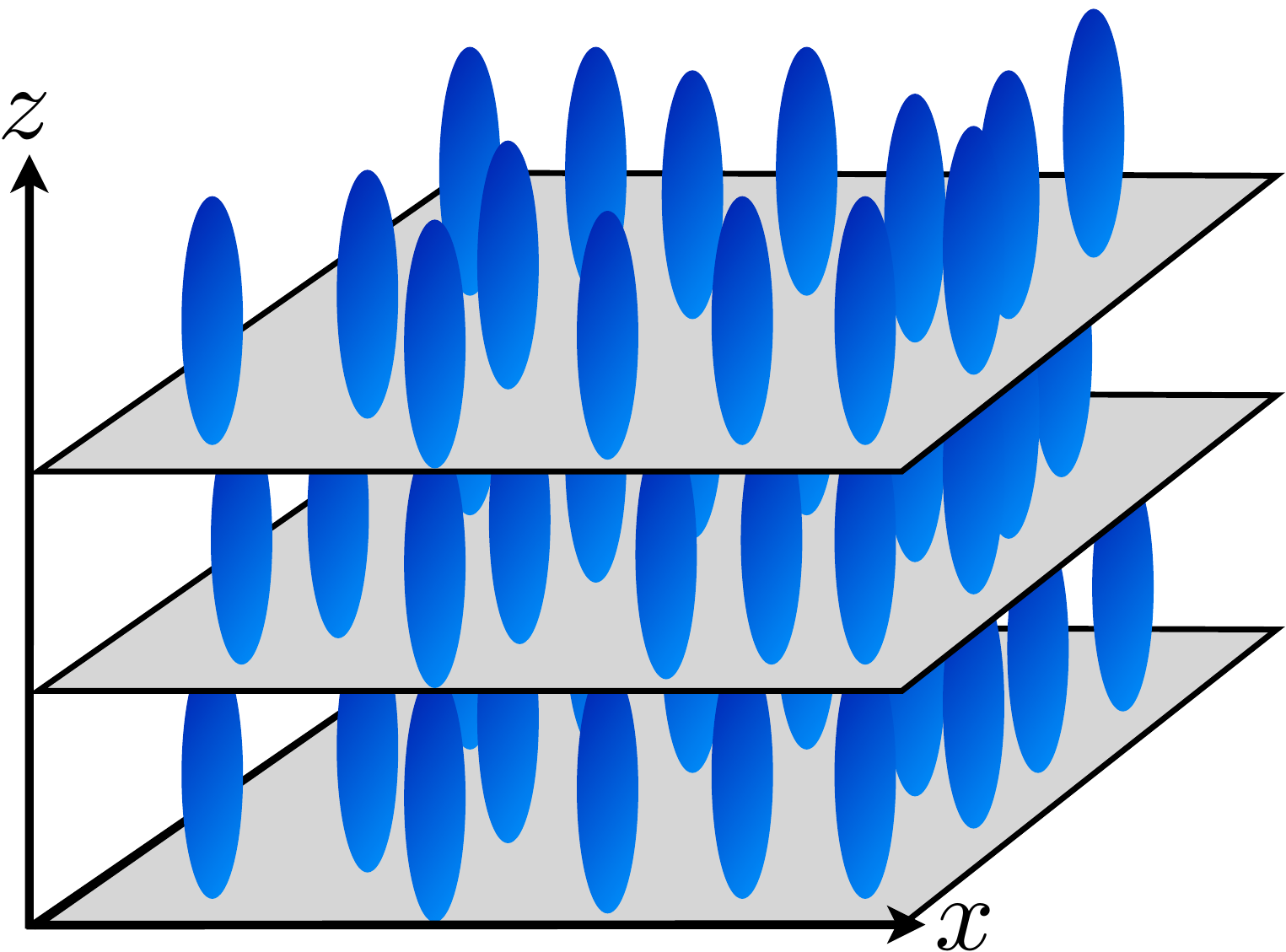}\\

 \vspace{-.4cm}

 \quad\hspace{1.cm}(a) \quad\hspace{3.4cm}(b)
\caption{\label{fig:LiquidCrystal}
Schematic figures for (a) nematic and (b) smectic-A  phases in a liquid crystal system.
}
\end{figure}
The first example is a liquid crystal system,
whose microscopic theory is invariant under rotational and translational transformations.
We consider the nematic and smectic-A phases of the liquid crystal, shown in Fig.~\ref{fig:LiquidCrystal}.
In both phases, the space-rotational symmetry is spontaneously broken, $O(3)\to O(2)$,
whose broken generators are $L_{xz}$ and $L_{yz}$.
The order parameter transforms under Eq.~(\ref{eq:chargeCommutation}) as
\begin{equation}
\begin{split}
&\average{[\hT_{\bm{R}}\hL_{iz}\hT_{\bm{R}}^{\dag},\hphi_k(\bm{x})]} \\
&\quad= \average{[\hL_{iz},\hphi_k(\bm{x})]}
- R_i\average{[\hP_z,\hphi_k(\bm{x})]}+R_z\average{[\hP_i,\hphi_k(\bm{x})]}.
\label{eq:transLij}
\end{split}
\end{equation}
In the nematic phase (Fig.~\ref{fig:LiquidCrystal}a), where the translation symmetry is not broken, the second and third terms in the right hand side of Eq.~(\ref{eq:transLij}) vanish, and thus ${A_a}^b(\bm{R})$=0. 
In this case, $\NEV=\NBS$, and 
two non-translationally invariant charges $\hL_{xz}$ and $\hL_{yz}$ generate two elastic variables.
In contrast, in the smectic-A phase (Fig.~\ref{fig:LiquidCrystal}b), the continuum translational symmetry along the $z$ direction is broken into the discrete one in addition to the rotational symmetry breaking, i.e., $\NBS=3$.
In this instance,  ${\AM_{P_{z}}}^{P_z}(\bm{R})={\AM_{P_{z}}}^{L_{iz}}(\bm{R})=0$ and ${\AM_{L_{iz}}}^{P_{z}}(\bm{R})=R_i$.
For each $R_i$, the dimension of $\AM$ is equal to two. However, one of them depends on $R_i$, while the other does not.
Thus, the dimension of $\AM$ for arbitrary $R_i$ is equal to one; there appears the only one elastic variable, which is associated with spontaneous breaking of translational symmetry~\cite{Chaikin}.

\subsection{Spontaneous breaking of conformal symmetry}
The second example is the system with conformal symmetry.
There are three types of non-translationally invariant charges $\hM_{\mu\nu}$, $\hD$, and $\hK_\mu$, which
are generators associated with Lorentz, scale, and special conformal transformations, respectively.
The commutation relations between these charges and translational operators are
\begin{align}
[\hM_{\mu\nu},\hP_\rho]&=-i(\eta_{\mu\rho}\hP_\nu-\eta_{\nu\rho}\hP_\mu),\\
[\hD,\hP_\mu]&=-i\hP_\mu,\\
[\hK_\mu,\hP_\nu]&=-2i(\eta_{\mu\nu}\hD+\hM_{\mu\nu}).
\end{align}
When a uniform condensation of a scalar field, $\average{\hat{\varPhi}(x)}\neq0$, exists, the scale and special conformal symmetries are spontaneously broken; $i\average{[\hD,\hat{\varPhi}(x)]}\neq0$ and $i\average{[\hK_\mu,\hat{\varPhi}(x)]}\neq0$. The number of broken symmetries is  five in ($3$+$1$) dimensions.
In this case, the system is still Lorentz invariant. Thanks to it, 
the elastic variables coincide the NG fields.
The matrix elements are ${\AM_D}^D(\epsilon)={\AM_{K_\mu}}^{K_\nu}(\epsilon)={\AM_D}^{K_\mu}(\epsilon)=0$, and ${\AM_{K_\mu}}^D(\epsilon)=2\epsilon_\mu$,
so that $ \mathop{\mathrm{dim}}\mathop{\mathrm{ker}}\AM=1$.  Thus, the only one NG mode appears, which is associated with broken scale invariance.
The NG mode corresponding to the special conformal symmetry does not appear~\cite{Volkov:1973vd, Ogievetsky,Low:2001bw,Isham:1970gz,ColemanText,  Higashijima:1994zg}.

\subsection{Quantum electrodynamics}
The last example is quantum electrodynamics, in which
the photons can be understood as  NG modes
 in a covariant gauge at zero temperature~\cite{Ferrari:1971at,Brandt:1974jw,Kugo:1985jc}.
We choose the gauge-fixing term as $\mathcal{L}_\text{GF}=B(x)\partial^\mu A_\mu(x)+\alpha (B(x))^2/2$, where $B(x)$, $A_\mu(x)$, and $\alpha$ are the Nakanishi-Lautrup field, the gauge fields, and the gauge parameter, respectively.
In this framework, again thanks to the Lorentz symmetry, the elastic variables are identical to the NG fields.
In this gauge fixing condition, there are two charges associated with residual gauge symmetries, $\delta A_\mu(x) = \partial_\mu\theta(x)$ with $\theta(x)=a_\mu x^\mu + b$:
$\hQ_\mu$ and $\hQ$. The commutation relations between these charges and the translational operators satisfy
 $[\hP_\nu, \hQ]=0$ and $[\hP_\nu,\hQ_\mu]=i\eta_{\nu\mu}\hQ$. Therefore,  $\hQ_\mu$ is non-translationally invariant charge.
Furthermore, $\hQ_\mu$ is always broken since the residual gauge transformation gives $\delta \hA_\nu= i[a^{\mu}\hQ_\mu+b\hQ, \hA_{\nu}(x)]=a_\nu$,
so that $i\average{ [\hQ_\mu, \hA_{\nu}(x)]}=\eta_{\mu\nu}$.
If $\hQ$ is not broken,  $\mathop{\mathrm{dim}}\mathop{\mathrm{ker}} A=4$.
Therefore, there appear four NG modes:
Two transverse components of them correspond to the physical photons, while the longitudinal and scalar components correspond to  unphysical NG modes that do not appear in the physical state. 
On the other hand, if $\hQ$ is spontaneously broken, i.e., if the vacuum is in the Higgs phase, 
$\NBS=5$,  ${\AM_{Q}}^{Q}(\epsilon)={\AM_{Q}}^{Q_\mu}(\epsilon)=0$ and  ${\AM_{Q_\mu}}^{Q}(\epsilon)=\epsilon_\mu$. We have $\mathop{\mathrm{dim}}\mathop{\mathrm{ker}}\AM=1$.
 This fact implies that the photons are no longer NG modes nor massless~\cite{Ferrari:1971at,Brandt:1974jw,Kugo:1985jc}.
In addition, the NG mode associated with broken $\hQ$ becomes an unphysical mode.
We note that this result depends on the choice of gauge fixing condition. For example, if one chooses $R_\xi$ gauge, which explicitly breaks the global $U(1)$ symmetry, 
there is no massless NG mode, even in the unphysical sector, although the physical spectra are independent from the choice of gauge fixing condition.
\section{Summary and discussion}
 We have discussed the relation between elastic variables and spontaneous breaking of continuum symmetries including spacetime ones at finite temperature.
 The elastic variables are given as the degrees of freedom corresponding to the flat directions of the free energy.
The general counting rule of elastic variables is given by Eq.~(\ref{eq:CountingElasticVariable}).
Although we begun with the free energy with a set of local fields at finite temperature, 
the result does not depend on the choice of local fields, and only depends on the symmetry breaking pattern.
Our result also works at zero temperature by taking the $T\to0$ limit . 

For symmetry breaking, whose generator explicitly depends on the time coordinate, i.e., $[\hQ_A, \hH]\neq0$, $\hQ_A$ does not generate an elastic variable at finite temperature,
because $[\hrho_\text{eq},Q_A]\neq0$, and thus Eq.~(\ref{eq:zeroEigenvalue}) is not satisfied. This is the case for Lorentz or Galilean boost,
in which the order parameter is $i\average{ [\hM_{0i}, \hT_{0j}(\bm{x})] }=\eta_{ij}h$ with the enthalpy $h$.
In fact, the susceptibility of $\hT_{0i}(\bm{x})$ is proportional to $h$ and does not diverge~\cite{Minami:2012hs}. 
Note, however, that $T_{0i}(\bm{x})$ contains the acoustic phonon as the NG mode associated with broken boost symmetry,
although $T_{0i}(\bm{x})$ is not elastic variable~\cite{Schakel:1996,hidaka}.

In this paper, we only focused on the elastic variables, not NG modes, which should be distinguished.
As is discussed in Refs.~\cite{Watanabe:2012hr,Hidaka:2012ym,Nambu:2004}, the elastic variables are not always independent in the sense of
the canonical variables. This will also happen for spacetime symmetries if $\average{[\hQ_a,\hQ_b]}\neq0$. 

It should be remarked on the previous work by Low and Manohar~\cite{Low:2001bw},
who discussed NG modes rather than elastic variables in Lorentz invariant theories at zero temperature.
In their analysis, the fluctuation field is introduced as 
$\delta\phi_i(\bm{x}) \equiv c^a(\bm{x}){[T_a]_i}^j\average{\hphi_j(\bm{x})}$ with $\average{[\hQ_a,\hphi_i(\bm{x})]}={[T_a]_i}^j\average{\hphi_j(\bm{x})}$.
The nontrivial solution satisfying 
\begin{equation}
iP_k\delta\phi_i(\bm{x}) = (\partial_k c^a(\bm{x})-c^b(\bm{x}) {f_{k b}}^a){[T_a]_i}^j\average{\hphi_j(\bm{x})} =0 \label{eq:LowManohaCondition}
\end{equation}
does not generate gapless excitations~\cite{Low:2001bw,Ivanov:1975zq}. 
Their result for constant $c^a$ reduces to our result for the elastic variable, when the unbroken translational symmetry is continuum.
We emphasize that their counting rule for NG modes does not work at finite temperature and/or non-Lorentz invariant systems, in particular,
when $\average{[\hQ_a,\hQ_b]}\neq0$.

For  dispersions of NG modes associated with spontaneous breaking of  spacetime symmetries,
they will not be robust; type-I and type-II do not coincide with type-A and type-B.
For example, NG modes in a nematic phase of liquid crystal, which are  type-A NG modes in the sense of $\average{[\hQ_a,\hQ_b]}=0$,
have the dispersions, $\omega= ak^2+ibk^2$ with constants $a$ and $b$~\cite{Hosino:1982}. The real and imaginary parts have the same order in $k$,
while for spontaneous breaking of internal symmetries, the imaginary part of dispersion of the NG mode is smaller than the real part at small $k$~\cite{Chaikin,hidaka}.
This behavior can be understood as follows:
The NG mode can be regarded as a propagating wave in which the elastic variable and the charge density are the canonical pair.
Their equations of motion (the generalized Langevin equations) are formally derived from Mori's projection operator method~\cite{Hidaka:2012ym,Mori},
where the expectation value of commutation relation between operators plays a role of Poisson bracket. 
Since the charge density of the rotational symmetry $\hat{L^0}_{ij}(x) = x_i\hT^0_{~j}(x)-x_j\hT^0_{~i}(x)$ explicitly depends on the spatial coordinates $x_i$ that are not operators,
one should employ $\hT^0_{~i}(x)$ as the pair operator of the elastic variable rather than $\hat{L}^0_{~ij}(x)$.
When only the rotational symmetry is spontaneously broken, the expectation value of commutation relation between the elastic variable and $\hat{L}^0_{ij}(x)$ does not vanish at momentum $k=0$, 
while that between the elastic variable and $\hT^0_{~j}(x)$ does.
At small momentum, it is proportional to $k$ because  $x_i$ in $\hat{L^{0}}_{ij}(x)$ turns to $\partial_{k_i}$ in momentum space.
Therefore, the additional $k$ appears in the Poisson bracket between the elastic variable and $\hT^0_{~i}(x)$, which makes the dispersion of NG mode quadratic~\cite{hidaka}. 
This suggests that the power of $k$ in the real part of the dispersion relation is strongly related to the power of $x$ in the nontranslationally invariant charge density. In contrast,  the imaginary part comes from the diffusion term in the Langevin equation for the charge density, which is obtained from the Kubo formula
that does not contain the Poisson bracket. Thus, the power of $k$ in the imaginary part for a type-A NG mode is the same as that for internal symmetry breaking, i.e., the order of $k^2$~\cite{hidaka}.  Note that for a type-B NG mode, the situation is different because the charge density itself is the elastic variable. For the internal symmetry breaking, the imaginary part behaves like $k^4$~\cite{hidaka}.
In general, the parameter $a$ strongly depends on temperatures; at some temperature $a$ vanishes, 
and thus, the mode is not propagating but overdamping~\cite{deGennesText}. 

Another nontrivial example is a capillary wave or ripplon, which propagates along the phase boundary of a fluid, and has the fractional dispersion, $\omega \sim k^{3/2}$~\cite{Landau:Fluid,Takeuchi:2013mwa,Watanabe:2014zza}. This dispersion does not belong either type-I or type-II NG modes in the Nielsen-Chadha classification~\cite{Nielsen:1975hm}.
In this case, the translation normal to the phase boundary of the fluid $T^{0z}$ is spontaneously broken, and  the mass density $\rho$ becomes the elastic variable. 
Here, we chose $z$ axis as the normal direction of the phase boundary.
Since the expectation value of the translation and the mass density is nonzero, the capillary wave is classified as a type-B NG mode.  For internal symmetry breaking, the time derivative of charge density behaves like $\partial_0 n_a\sim \partial_i^2 n_b$, where $n_a$ and $n_b$ are broken charge densities~\cite{Watanabe:2012hr,Hidaka:2012ym}, and it leads to the quadratic dispersion $\omega^2 \sim k^4$. On the other hand, for the capillary wave, the time derivative of $\rho$ is given by 
its continuity equation as $\partial_0\rho=-\partial_i T^{0i}\sim -\partial_z T^{0z}$. 
The power of $k$ in the equation of motion for $\rho$ is reduced to linear from quadratic for a conventional type-B NG mode,
while the equation of motion for $T^{0z}$ is the same as the conventional one,  $\partial_0 T^{0z}\sim \partial_i^2 \rho$ ($i=x,y$).
As a result, we have $\omega^2\sim k^3$, and thus, the fractal dispersion relation is realized $\omega\sim k^{3/2}$.
It is interesting future problem to investigate the dispersion relations of NG modes associated with spontaneous breaking of spacetime symmetries.

\section*{ACKNOWLEDGMENTS}
We thank  Y.~Hirono, T.~Kugo, Y.~Tanizaki, and A.~Yamamoto for useful discussions.  
T.~H.~was supported by JSPS Research Fellowships for Young Scientists.
Y.~H. was partially supported by JSPS KAKENHI Grants Numbers 24740184,  23340067. 
This work was also partially supported  by RIKEN iTHES Project.

\bibliographystyle{elsarticle-num}
\bibliography{ngmode}

\providecommand{\noopsort}[1]{}\providecommand{\singleletter}[1]{#1}%
\begin{thebibliography}{10}
\expandafter\ifx\csname url\endcsname\relax
  \def\url#1{\texttt{#1}}\fi
\expandafter\ifx\csname urlprefix\endcsname\relax\def\urlprefix{URL }\fi
\expandafter\ifx\csname href\endcsname\relax
  \def\href#1#2{#2} \def\path#1{#1}\fi

\bibitem{Chaikin}
P.~M. Chaikin, T.~C. Lubensky, {Principles of Condensed Matter Physics},
  Cambridge University Press, 2000.

\bibitem{Coleman:1969sm}
S.~R. Coleman, J.~Wess, B.~Zumino, {Structure of phenomenological Lagrangians.
  1.}, Phys. Rev. 177 (1969) 2239--2247.

\bibitem{Callan:1969sn}
C.~G. Callan~Jr., S.~R. Coleman, J.~Wess, B.~Zumino, {Structure of
  phenomenological Lagrangians. 2.}, Phys. Rev. 177 (1969) 2247--2250.

\bibitem{Volkov:1973vd}
D.~V. Volkov, {Phenomenological Lagrangians}, Sov. J. Part. Nucl. 4 (1973) 3.

\bibitem{Ogievetsky}
V.~I. Ogievetsky, {Nonlinear Realizations of Internal and space-time
  symmetries}, Proc. of X-th Winter School of Theoretical Physics in Karpacz 1
  (1974) 117.

\bibitem{Nambu:1961tp}
Y.~Nambu, G.~Jona-Lasinio, {Dynamical Model of Elementary Particles Based on an
  Analogy with Superconductivity. 1.}, Phys. Rev. 122 (1961) 345--358.

\bibitem{Goldstone:1961eq}
J.~Goldstone, {Field Theories with Superconductor Solutions}, Nuovo Cim. 19
  (1961) 154--164.

\bibitem{Goldstone:1962es}
J.~Goldstone, A.~Salam, S.~Weinberg, {Broken Symmetries}, Phys. Rev. 127 (1962)
  965--970.

\bibitem{Gasser:1984}
J.~Gasser, H.~Leutwyler, Chiral perturbation theory to one loop, Annals of
  Physics 158~(1) (1984) 142--210.

\bibitem{Hofmann:1999}
C.~P. Hofmann, Effective analysis of the $\mathrm{O}(n)$
  antiferromagnet:\quad{}low-temperature expansion of the order parameter,
  Phys. Rev. B 60 (1999) 406--413.

\bibitem{Hofmann:2001ck}
C.~P. Hofmann, {Spontaneous magnetization of the O(3) ferromagnet at low
  temperatures}, Phys. Rev. B 65 (2002) 094430.
\newblock \href {http://arxiv.org/abs/cond-mat/0106492}
  {\path{arXiv:cond-mat/0106492}}.

\bibitem{Anderson:2002}
J.~O. Andersen, {Effective Field Theory for Goldstone Bosons in Nonrelativistic
  Superfluids}, arXiv:cond-mat/0209243.

\bibitem{Anderson:2004}
J.~O. Andersen, Theory of the weakly interacting bose gas, Rev. Mod. Phys. 76
  (2004) 599--639.

\bibitem{Brauner:2010wm}
T.~Brauner, {Spontaneous Symmetry Breaking and Nambu-Goldstone Bosons in
  Quantum Many-Body Systems}, Symmetry 2 (2010) 609--657.
\newblock \href {http://arxiv.org/abs/1001.5212} {\path{arXiv:1001.5212}}.

\bibitem{Watanabe:2012hr}
H.~Watanabe, H.~Murayama, {Unified Description of Nambu-Goldstone Bosons
  without Lorentz Invariance}, Phys. Rev. Lett. 108 (2012) 251602.
\newblock \href {http://arxiv.org/abs/1203.0609} {\path{arXiv:1203.0609}}.

\bibitem{Hidaka:2012ym}
Y.~Hidaka, {Counting rule for Nambu-Goldstone modes in nonrelativistic
  systems}, Phys. Rev. Lett. 110 (2013) 091601.
\newblock \href {http://arxiv.org/abs/1203.1494} {\path{arXiv:1203.1494}}.

\bibitem{Nambu:2004}
Y.~Nambu, Spontaneous breaking of lie and current algebras, Journal of
  Statistical Physics 115 (2004) 7--17.

\bibitem{Nielsen:1975hm}
H.~B. Nielsen, S.~Chadha, {On How to Count Goldstone Bosons}, Nucl. Phys. B105
  (1976) 445.

\bibitem{Low:2001bw}
I.~Low, A.~V. Manohar, {Spontaneously broken spacetime symmetries and
  Goldstone's theorem}, Phys. Rev. Lett. 88 (2002) 101602.
\newblock \href {http://arxiv.org/abs/hep-th/0110285}
  {\path{arXiv:hep-th/0110285}}.

\bibitem{Ivanov:1975zq}
E.~Ivanov, V.~Ogievetsky, {The Inverse Higgs Phenomenon in Nonlinear
  Realizations}, Teor. Mat. Fiz. 25 (1975) 164--177.

\bibitem{Nicolis:2013sga}
A.~Nicolis, R.~Penco, F.~Piazza, R.~A. Rosen, {More on gapped Goldstones at
  finite density: More gapped Goldstones}, JHEP 1311 (2013) 055.
\newblock \href {http://arxiv.org/abs/1306.1240} {\path{arXiv:1306.1240}}.

\bibitem{Endlich:2013vfa}
S.~Endlich, A.~Nicolis, R.~Penco, Phys. Rev. D 89 (2014) 065006.
\newblock \href {http://arxiv.org/abs/1311.6491} {\path{arXiv:1311.6491}}.

\bibitem{Brauner:2014aha}
T.~Brauner, H.~Watanabe, {Spontaneous breaking of spacetime symmetries and the
  inverse Higgs effect}, Phys. Rev. D89 (2014) 085004.
\newblock \href {http://arxiv.org/abs/1401.5596} {\path{arXiv:1401.5596}}.

\bibitem{Watanabe:2013iia}
H.~Watanabe, H.~Murayama, {Redundancies in Nambu-Goldstone Bosons}, Phys. Rev.
  Lett. 110 (2013) 181601.
\newblock \href {http://arxiv.org/abs/1302.4800} {\path{arXiv:1302.4800}}.

\bibitem{WeinbergText}
S.~Weinberg, {The Quantum Theory of Fields, Vol. II}, Cambridge University
  Press, Cambridge, UK, 1996.

\bibitem{Isham:1970gz}
C.~Isham, A.~Salam, J.~Strathdee, {Spontaneous breakdown of conformal
  symmetry}, Phys. Lett. B 31 (1970) 300--302.

\bibitem{ColemanText}
S.~Coleman, {Aspects of Symmetry}, Cambridge University Press, Cambridge, UK,
  1985.

\bibitem{Higashijima:1994zg}
K.~Higashijima, {Nambu-Goldstone theorem for conformal symmetry}, In
  Proceedings of XX International Colloquium on Group Theoretical Methods in
  Physics (1994) 223--228.

\bibitem{Ferrari:1971at}
R.~Ferrari, L.~Picasso, {Spontaneous breakdown in quantum electrodynamics},
  Nucl. Phys. B31 (1971) 316--330.

\bibitem{Brandt:1974jw}
R.~A. Brandt, W.-C. Ng, {Gauge Invariance and Mass}, Phys. Rev. D 10 (1974)
  4198.

\bibitem{Kugo:1985jc}
T.~Kugo, H.~Terao, S.~Uehara, {Dynamical Gauge Bosons and Hidden Local
  Symmetries}, Prog. Theor. Phys. Suppl. 85 (1985) 122--135.

\bibitem{Minami:2012hs}
Y.~Minami, Y.~Hidaka, {Relativistic hydrodynamics from the projection operator
  method}, Phys. Rev. E 87~(2) (2013) 023007.
\newblock \href {http://arxiv.org/abs/1210.1313} {\path{arXiv:1210.1313}}.

\bibitem{Schakel:1996}
A.~M.~J. Schakel, {Effective Field Theory of ideal-fluid Hydrodynamics}, Mod.
  Phys. Lett. B 10 (1996) 999.
\newblock \href {http://arxiv.org/abs/cond-mat/9607164}
  {\path{arXiv:cond-mat/9607164}}.

\bibitem{hidaka}
T.~Hayata, Y.~Hidaka, Y.~Hirono, in preparation.

\bibitem{Hosino:1982}
M.~Hosino, H.~Nakano, {Molecular Theory of Hydrodynamic Equations for Nematic
  Liquid Crystals}, Prog. Theor. Phys. 68 (1982) 388--401.

\bibitem{Mori}
H.~Mori, Transport, collective motion, and brownian motion, Prog. Theor. Phys.
  33~(3) (1965) 423--455.

\bibitem{deGennesText}
P.~G. de~Gennes, P.~J., {The Physics of Liquid Crystals}, Oxford University
  Press, 1993.

\bibitem{Landau:Fluid}
L.~D. Landau, E.~M. Lifshitz, {Fluid Mechanics, Second Edition}, Butterworth
  Heinemann, Oxford, UK, 1987.

\bibitem{Takeuchi:2013mwa}
H.~Takeuchi, K.~Kasamatsu, Nambu-goldstone modes in segregated bose-einstein
  condensates, Phys. Rev. A 88 (2013) 043612.
\newblock \href {http://arxiv.org/abs/1309.3224} {\path{arXiv:1309.3224}}.

\bibitem{Watanabe:2014zza}
H.~Watanabe, H.~Murayama, {Nambu-Goldstone bosons with fractional-power
  dispersion relations}, Phys. Rev. D89 (2014) 101701.
\newblock \href {http://arxiv.org/abs/1403.3365} {\path{arXiv:1403.3365}}.

\end{thebibliography}
\end{document}